\documentclass[twocolumn,preprintnumbers,amsmath,amssymb]{revtex4}
\usepackage[dvipdfmx]{graphicx}
\usepackage{dcolumn}
\usepackage{bm}
\usepackage{color}
\usepackage{comment}
\begin{document}

\title{Self-replicating segregation patterns in horizontally vibrated binary mixture of granules}
\author{Hiroyuki Ebata}
\email{ebata.hiroyuki@phys.kyushu-u.ac.jp}
\author{Shio Inagaki}
\affiliation{%
Graduate School of Science, Kyushu University, 744 Motooka, Nishi-ku, Fukuoka 819-0395, Japan
}
\date{\today}

\begin{abstract}
When granular mixtures of different sizes are fluidized, each species spontaneously separates and condenses to form patterns.   
Although granular segregation has been extensively studied, the inability to directly observe the time evolution of the internal structure hinders the understanding of the mechanism of segregation dynamics driven by surface flow. 
In this study, we report rich band dynamics, including a self-replicating band, in a horizontally shaken granular mixture in a quasi-two-dimensional container where the granules formed steady surface waves. 
Direct observation of surface flow and segregated internal structure revealed that coupling among segregation, surface flow, and hysteresis in the fluidity of granules is key to understanding complex band dynamics. 
\end{abstract}

\maketitle
\;
{\bf Introduction.}
\;
Granular materials composed of solid macroscopic particles are widely used in both natural and industrial processes.
In the industry, the considerable need for efficient handling of granular materials has been the focus of research \cite{Williams76:review, Zafar2017:review}.
Granular materials are a typical example of a system far from equilibrium.
When mechanical agitation, including flow, vibration, and rotation, is applied, a balance between energy injection and dissipation owing to frictional and inelastic interactions leads to various dissipative structures \cite{Jaeger96:review, deGennes1999:review, Aranson:Pattern}.
\\ \;
One of the most counterintuitive phenomena is segregation, which is caused by the differences in size, density, or other physical properties of the particles \cite{Breu03:_BrazilNuts, Xu2017:_BrazilNuts, Rehberg:Swirling}.
For example, the alternating layer emerges spontaneously in a quasi-two-dimensional sandpile as a binary granular mixture is poured between two vertical plates, which is called stratification \cite{Makse:Stratification}.
In the rotating cylinder, alternating bands appear and are subsequently merged as the rotation proceeded 
\cite{Ottino00:_SegregationRevieiw,Thomas00:_SegregationRevieiw}.
When heavy particles are placed on top of a layer of light particles on a slope, the upper layer of heavy particles splits and forms strips parallel to the direction of the chute flow \cite{Dortona20:_RT}. The strips merge as they flow downward as the axial bands merge in a rotating cylinder.
\begin{figure*}[btp]
    \begin{center}
    \includegraphics[width=16cm]{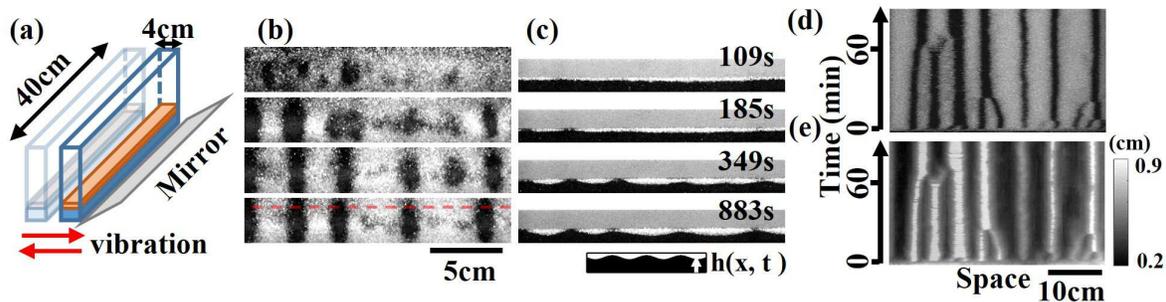}
    \caption{
    (a) Schematic illustration of the experimental system.
    (b, c) Time series of upper (b) and side (c) view of the band formation. As an initial condition, two granular species are uniformly mixed. We only showed a part of image (18 cm out of 40 cm in length). 
    (d, e) Spatiotemporal plot of the bands (d) and height $h(x,t)$ between bead and frit layers (e). (d) The image pixels of the red dashed line in Fig. \ref{fig:fig1} (b) were stacked. (e) The color indicates the height (see colorbar).
     (b - e) Width of the container is 3.5 cm. Diameter of glass frits and beads are $D_{f} =$ 381 $\mu$m and $D_{b} =$ 868 $\mu$m, respectively. Vibration frequency $f$ = 2.8 Hz. (d, e) Width and height of the images are 40 cm and 5450 s, respectively.
    }
    \label{fig:fig1}
    \end{center}
\end{figure*}
  \begin{figure}[tbp]
      \begin{center}
      \includegraphics[width=7cm]{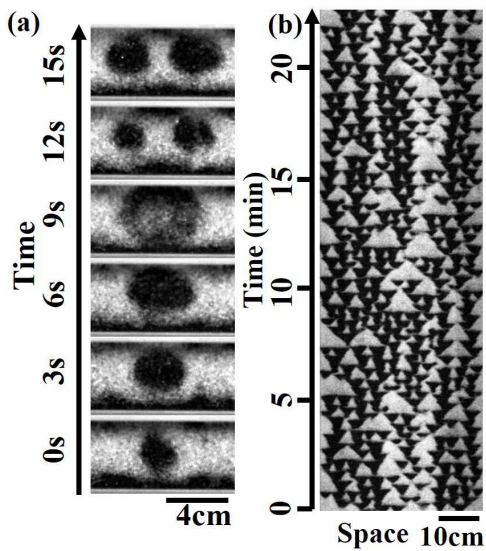}
      \caption{
        Time series (a) and spatiotemporal plot (b) of the replicating bands. (a, b) The granules were initially segregated in the depth direction. Width of the container was 4 cm. $D_{f} =$ 337 $\mu$m. $D_{b} =$ 1240 $\mu$m. $f$ = 3.0 Hz (b) Width and height of the image are 40 cm and 1340 s, respectively.
      }
      \label{fig:fig2}
      \end{center}
  \end{figure}
\\ \;
Although the aforementioned segregations were driven by surface granular flow, it remains unclear whether physical properties such as the rheology of granules or hydrodynamic instability are the dominant driving mechanisms.
In the case of axial segregation in a rotating cylinder, the difference in the dynamic angles of repose, which relates to the fluidity of granules, has been considered a macroscopic parameter that determines segregation. However, several studies claim that the axial band formation may be due to the bulging of the radial core of small particles \cite{Alexander04:_Segregation,Balista:_AxialCore}. Alternatively, the external layer of large particles with high density may be destabilized by Rayleigh-Taylor like instability \cite{Dortona20:_RT}.
However, it is difficult to answer this question as it is not possible to observe the time evolution of the internal structure with sufficiently high resolutions both in time and space in a non-invasive manner.
\\ \;
In this study, we conducted experiments using a mixture of spherical and angular particles in a horizontally shaken narrow channel.
This enabled us to measure the time evolution of the surface and internal structures simultaneously.
The surface flow driven by horizontal vibration induced segregation bands even without a density difference, whose onset was reminiscent of fluid-fluid interface instability. 
In addition to the coarsening patterns commonly observed in rotating cylinder systems,
a nonstationary state was observed: a self-replicating segregation pattern. 
Measurement of the fluidity of each granular species and the phenomenological model of pattern formation suggested that the interplay among segregation, surface flow, and granular fluidity with strong hysteresis is key to understanding complex and rich band dynamics. 
\\ \;
{\bf Experimental conditions.}
\;
To investigate how the surface flow induces the segregation pattern, we applied horizontal vibration to a granular mixture bed (depth $\sim$ 1 cm) in a rectangular container [Fig. \ref{fig:fig1} (a)]. The length and width of the container were 40 and 4 cm, respectively, unless otherwise stated.
Glass beads (diameter 0.8 mm) were glued to the bottom of the container to prevent slipping of the granular bed. 
In this study, the direction of vibration was parallel to the short axis of the container, and the amplitude of the vibration (5 cm peak-to-peak) was larger than that of the short axis.
We varied the vibration frequency, $f$ from 0 to 3.0 Hz. 
\\ \;
We used a mixture of glass beads (white color, 2.56 g / cm$^3$) and glass frits (black color, 2.57 g / cm$^3$) to exclude the effect of the density difference. 
Since the glass frits had a wide distribution of particle sizes, we sieved the glass frits into several average sizes $D_{f}$ ranging from 150 to 650 $\mu$m (see supplementary information). 
The size $D_{b}$ of the glass beads ranged from 400 to 2000 $\mu$m. 
The glass frits were elongated polygons. 
Therefore, compared with glass beads, glass frits require a higher vibration frequency to be fluidized. 
For the initial condition of the granular beds, a layer of large granules was placed on the layer of small granules, unless otherwise stated. 
This initially segregated condition can achieve high reproducibility of the experimental results. 
\\ \; Using a CCD camera, we observed the top and side views of the container simultaneously.
From the side view, we measured the segregation in the depth direction.  
The image was obtained at a certain vibration phase by detecting the container position with the laser sensor. 
Using a high-speed camera (HAS U-1, Detect), we recorded a video of the lateral view of the granular bed to analyze the surface wave.
\\ \;
{\bf Band formation driven by surface waves.}
\;
When uniformly mixed glass beads and frits were horizontally vibrated at a sufficiently large frequency, the fluidized granules formed a surface wave whose propagation direction was parallel to the short axis of the container. 
The surface wave drove the segregation in the vertical and horizontal directions simultaneously (Fig. \ref{fig:fig1} (b, c)).
The time evolution resembled radial and axial segregation in a rotating drum \cite{Ottino00:_SegregationRevieiw,Thomas00:_SegregationRevieiw, Nitya:RadSeg, Zik94:Segregation}. 
First, large glass beads (white) covered the surface of the granular beds. Bands of glass frits (black) then appeared and merged to form wider bands, as shown in the spatiotemporal plot (Fig. \ref{fig:fig1} (d)).
From the side view of the container (Fig. \ref{fig:fig1} (c)), band formation appeared as an interface instability such as Rayleigh Taylor Instability \cite{Taylor:Instab, Dortona20:_RT}; surface glass beads' layer split to form ``droplets'' as the undulation of the interface between the bead and frit layers grew and reached the surface.
A spatiotemporal plot of the band (Fig. \ref{fig:fig1} (d)) corresponded to that of the interface height $h(x,t)$ (Fig. \ref{fig:fig1} (e)). 
\\ \;
{\bf Self-replicating bands.}
\;
In contrast to axial segregation in the rotating drum, complex band dynamics, i.e., self-replication, emerged with a rather large particle size ratio $\rho_{ls} \sim 3$. The initial small band gradually increased in size and eventually split into two bands [Fig. \ref{fig:fig2} (a)]. 
The band has a typical size above which it becomes unstable and starts to replicate. 
When two or more bands collide, annihilation occurs. 
The replication cycle ($\sim$ 15 s) was considerably slower than the vibration cycle ($\sim$ 0.33 s). 
The dynamics of replicating bands resemble those of replicating patterns observed in reaction-diffusion systems \cite{Lee:selfRep, Peason:selfRep} and vertically vibrated suspensions \cite{Ebata:RepholePRL}. 
\\ \;
Owing to the successive replication and annihilation of the bands, the spatiotemporal plot of the replicating bands showed a Sierpinski gasket-like pattern (Fig. \ref{fig:fig2} (b)) found in the replicating patterns in 1-dimensional reaction-diffusion systems \cite{Hayase:SKgasketPRL, Hayase:SKgasketPRE}. 
The small black triangles in the plot are derived from a cycle of band replication, whereas the large white triangles are obtained from the pair and multi-annihilation of the bands. 
In contrast to the spatially symmetric patterns of the Sierpinski gasket, the spatiotemporal pattern of the replicating bands was disturbed by partial replication \cite{Ebata:RepholePRE}, partial annihilation \cite{Ebata:RepholePRL, Descalzi:PartAnni}, spontaneous creation, and spontaneous annihilation of the bands. 
    \begin{figure}[tbp]
      \begin{center}
      \includegraphics[width=7.5cm]{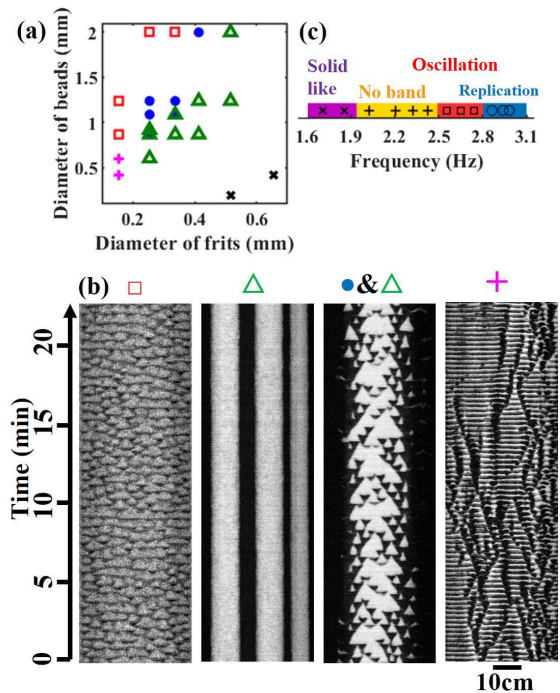}
      \caption{
      (a) Phase diagram of the pattern of the combination of particle sizes, $D_{f}$ and $D_{b}$. $f$ = 3.0 Hz (b) Typical spatiotemporal plots of the band dynamics.
      (a, b)
      Red square: oscillation.
      Green triangle: coarsening. 
      Blue circle: replication. 
      Pink cross: traveling wave.
      Black cross: no bands. 
      Width and height of the image correspond to 40 cm and 1340 s, respectively.
      (c) Dependence of band dynamics on vibration frequency $f$. $D_{f} =$ 337 $\mu$m. $D_{b} =$ 1240 $\mu$m.
      }
      \label{fig:fig3}
      \end{center}
  \end{figure}
\\ \;
{\bf Phase diagram of band dynamics.}
In addition to replication, the horizontally vibrated granular mixture exhibited various band dynamics depending on the combination of particle sizes (Figs. \ref{fig:fig3} (a) and (b)). 
In the case of a small size ratio ($\rho_{ls} < 3$), the band dynamics showed a coarsening process \cite{Mullin00:SegrelationPlate, Frette:SegCorsening}; the initial small bands merged to create larger bands (Figs \ref{fig:fig1} (b) and (d)). 
After coarsening, several bands remained, reaching a steady state. 
As the particle size ratio increased, the stationary bands bifurcated into oscillatory patterns through the replicating bands. 
The two types of band dynamics coexist in the marginal parameter between the stationary and replicating bands. 
When the glass beads and frits were both small, the system was disturbed. 
The traveling wave of glass beads and frits alternately propagated from the longer side of the container to the shorter side, resulting in the oscillatory pattern in the spatiotemporal plot [pink cross in Fig. \ref{fig:fig3} (a, b)]. 
Band formation did not occur for $\rho_{ls} < 1$. Thus, for band formation, the size of the glass beads should be larger than that of the frits. 
This suggests that the layer of large particles should be easily fluidized to form a surface wave, which drives the band formation. 
\\ \;
To determine how the magnitude of the surface wave alters the band dynamics, we investigated the dependence of the band pattern on vibration frequency (Figure \ref{fig:fig3} (c)). 
When the vibration frequency was sufficiently small, the granular bed was stationary and the surface flow was not induced. 
As the vibration frequency increased, large particles began to flow; however, band formation did not occur. 
Above a certain critical frequency, small bands appeared and disappeared in an oscillatory manner. 
When the vibration frequency increased further, the bands began to replicate. 
Thus, the dynamics of surface flow, which reflects the rheology of shaken dry granules \cite{Marchal:GraRheo,Dijksman:GraRheoHys}, is key to understanding the mechanism of pattern formation. 
\begin{figure}[tbp]
\begin{center}
\includegraphics[width=7cm]{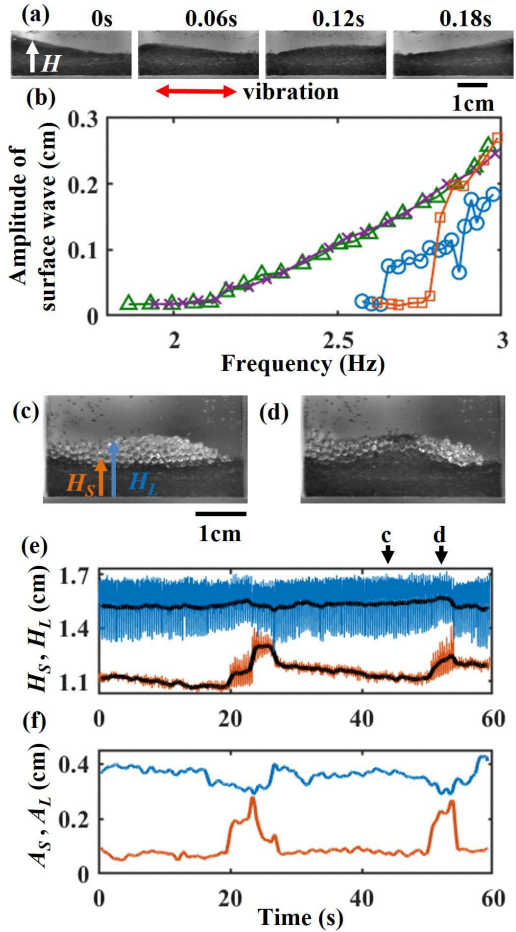}
\caption{
Dynamics of the surface wave driven by horizontal vibration. 
(a) Time series of the surface traveling wave of a glass frits' bed during a vibration cycle. Images were recorded from the lateral direction of the container. 
(b) Dependence of the amplitude of the surface wave on vibration frequency. Green triangle: beads' bed with decreasing frequency. Purple cross: beads' bed with increasing frequency. Blue circle: frits' bed with decreasing frequency. Red square: frits' bed with increasing frequency.
(c, d) Typical lateral view of the segregated granular mixture without band (c) and with band (d). 
$H_{L}$: height of the surface of the entire granular bed. 
$H_{S}$: height of the interface between the beads' and frits' layers. 
(e) Time evolution of $H_{L}$ and $H_{S}$. The replicating band appeared occasionally. 
(f) Time evolution of the oscillation amplitude $A_{L}$ and $A_{S}$ of $H_{L}$ and $H_{S}$. 
(a - f) $D_{f} =$ 337 $\mu$m. (b - f) $D_{b} =$ 1240 $\mu$m. (c - f) $f$ = 2.9 Hz.
}
\label{fig:fig4}
\end{center}
\end{figure}
\\ \;
{\bf Surface wave of granular bed.}
To characterize the surface flow of the granular bed, we first measured the amplitude of the surface wave of the granular bed with a single species (Fig. \ref{fig:fig4} (a)). 
Here, we observed the granular bed from the lateral view of the container parallel to the vibration direction and measured the heights of the surface $H$ (Fig. \ref{fig:fig4} (a)). 
We define the standard deviations of $H$ during one vibration cycle as the wave amplitudes. 
Figure \ref{fig:fig4} (b) shows the frequency ramp test; a large vibration frequency was initially imposed, and we gradually decreased the frequency until the surface flow was stopped. The frequency was then gradually increased. 
In the case of the glass beads, the layer of beads started to flow at a small vibration frequency. 
The surface wave amplitude did not exhibit a clear hysteresis for frequency as the glass beads easily rolled on the granular bed surface. 
In the case of glass frits, the layer of frits required a higher vibration frequency to flow, and the surface wave amplitude demonstrated a strong hysteresis. 
Even at a sufficiently large vibration frequency, surface waves have branches with higher and lower amplitudes, and such hysteresis of the flow has been found in frictional grains \cite{Dijksman:GraRheoHys, DeGiuli:GraflowHys, Perrin:GraflowHys}. 
When we observed the surface flow from above, the surface wave of the glass frits was spatially heterogeneous, and slower and faster-flowing regions coexisted in the container. 
Figures \ref{fig:fig3} (c) and \ref{fig:fig4} (b) indicate that the band appeared when the glass frits started to flow.  
\\ \;
To reveal how the surface flow interacted with the band dynamics, we measured the surface wave dynamics during the band formation of the granular mixture, where beads and frits were vertically segregated. 
Here, we calculated the surface height $H_{L}$ of the bead layer and the interface height $H_{S}$ between the bead and frit layers (Figs. \ref{fig:fig4} (c, d)). 
Figure \ref{fig:fig4} (e) shows the time series of $H_{L}$ (blue line), $H_{S}$ (orange line), and their time-averaged values (black lines) for replicating bands. 
A large oscillation of $H_{L}$ represents surface wave propagation from side to side (Fig. \ref{fig:fig4} (a)). 
As the surface was mainly covered by glass beads, $H_{L}$ was almost always larger than $H_{S}$. 
When a band appeared near the wall, $H_{S}$ sharply increased and approached $H_{L}$ (Figs. \ref{fig:fig4} (e)). 
During the appearance of the band, the magnitude of the oscillation of $H_{L}$ and $H_{S}$ changed, suggesting that the band altered the surface flow. 
\\ \;
To quantify the magnitude of the surface flow, we calculated the amplitudes of the waves, $A_{L}$ and $A_{S}$, which were defined as the standard deviations of $H_{L}$ and $H_{S}$ with one vibration cycle. 
As shown in Fig. \ref{fig:fig4} (f), when the band appeared near the wall, $A_{L}$ slightly decreased and $A_{S}$ sharply increased like a pulse in excitable media \cite{Meron:excitable}. 
The large difference between $A_{L}$ and $A_{S}$ should cause a shear stress at the interface. 
Thus, the flow of the glass frits must be agitated by the shear flow of glass beads. 
However, increasing the number of angular frits in the bead layer can suppress the surface flow $A_{L}$ (Fig. \ref{fig:fig4} (f) blue line), which could lead to the decline of the agitation of glass frits' layer. 
\\ \;
{\bf Discussion.} 
The experimental results suggest the following interplay between the surface wave and band dynamics. 
The shear flow at the interface between the glass beads and frits gradually rolled up the glass frits into a highly fluidized bead layer. 
The increasing number of frits in the bead layer leads to the growth of the band owing to segregation in the horizontal direction \cite{Zik94:Segregation, Mullin00:SegrelationPlate}. 
However, as Fig. \ref{fig:fig4} (b) implies, the highly fluidized state of glass frits was metastable \cite{Dijksman:GraRheoHys}. 
The condensation of the frits decreased the amplitude of the total surface wave owing to the low fluidity and triggered the jump into the lower fluidized state, which decreased $A_{s}$ and $A_{L}$. 
Consequently, the particle supply from the glass frits' layer decreased or stopped. 
Then, frits on the surface were absorbed into the basal frits layer owing to vertical segregation \cite{Nitya:RadSeg}, and a band of frits was annihilated. 
\begin{figure}[tbp]
\begin{center}
\includegraphics[width=8cm]{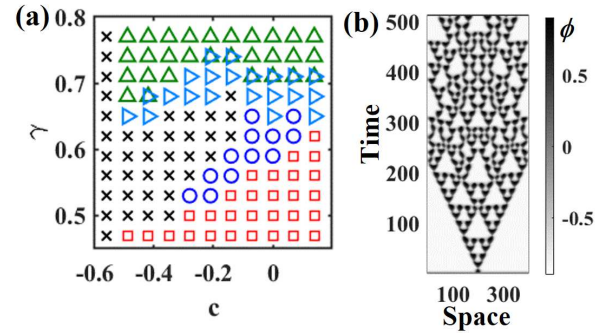}
\caption{
(a) Phase diagram of patterns obtained from Eqs. (\ref{eq:eq1}) and (\ref{eq:eq2}). 
Red square: oscillating pattern.
Green triangle: stable spot. 
Blue circle: replicating spot. 
Black cross: uniform state.
$\tau$ = 0.49, $d$ = 2.25, $\tau_{p}$ = 2.25, $I$ = 0.1, $a$ = 5, $D_{\phi}$ = 22.5, $D_{\psi}$ = 1.
(b) Typical spatiotemporal plot of the replicating spots.
}
\label{fig:fig5}
\end{center}
\end{figure}
\\ \;
Based on experimental observations, we proposed a phenomenological model of band dynamics. 
Here, we introduce two variables: the phase $\phi$ of the surface bead layer and the state of flow $\psi$ of the surface layer. 
We consider that $\phi$ increases as the number density of frits in the surface bead layer increases; $\phi < 0$ and $\phi > 0$ represent bead-rich and frits-rich phases, respectively.  
We also assume that $\psi$ is correlated with the magnitude of the surface flow; as $\psi$ increased, the amplitude of the surface wave increased. 
Thus, $\psi \ll 0$ indicates that the surface flow stops. 
\\ \;
Regarding the dynamics of $\phi$, $\phi$ was assumed to be linearly activated by $\psi$ because the small particles were blown up into the surface fluidized layer by surface flow. 
When the flow $\psi$ was constant, $\phi$ could relax to a certain value that is a function of $\psi$: Thus, the equation is
\begin{equation}
    \tau_{\phi}\frac{d \phi}{dt} = \psi - \gamma \phi + I + D_{\phi} \nabla^2 \phi, 
    \label{eq:eq1}
\end{equation}
where $\gamma$, $I$, and $D_{\phi}$ are the relaxation rate, offset (leak) of the phase, and diffusion coefficient, respectively.  
If the surface flow is weak ($\psi \ll 0$), a bead-rich phase ($\phi \ll 0$) is obtained in the surface layer, which corresponds to the no-band state in Fig. \ref{fig:fig2} (c). 
\\ \;
We considered that the flow $\psi$ depends on the phase $\phi$ of the surface fluidized layer because an increase in frits in the fluidized layer weakens the surface wave (Fig. \ref{fig:fig3} (f), blue line). 
Furthermore, $\psi$ can be bistable owing to the strong hysteresis of the frits flow \cite{Dijksman:GraRheoHys, Marchal:twostate} (Figs. \ref{fig:fig3} (b, e)). 
For simplicity, we introduce a cubic nonlinear term for the time-evolution equation of $\psi$ to implement bistability.
\begin{equation}
    \tau_{\psi}\frac{d \psi}{dt} = a\psi\left(1+\psi\right)\left(1-\psi\right) +c - d\phi + D_{\psi} \nabla^2 \psi, 
    \label{eq:eq2}
\end{equation}
where $c$ and $D_{\psi}$ are the offset and diffusion coefficients, respectively. 
Eqs. (\ref{eq:eq1}) and (\ref{eq:eq2}) are Bonhoeffer-van der Pol type reaction-diffusion equations, where self-replicating pulses have been found \cite{Hayase:SKgasketPRE}. 
\\ \;
Figure \ref{fig:fig5} shows the phase diagram of the model. As the relaxation rate $\gamma$ increases, the pattern bifurcates from an oscillation to a stable spot through replicating spots. 
A comparison of the experimental results (Fig. \ref{fig:fig3}) and the simulation (Fig. \ref{fig:fig5}) implies that relaxation rate $\gamma$ is a decreasing function of particle size ratio $\rho_{ls}$. 
Considering that $\gamma$ in the model describes how fast segregation proceeds, $\gamma$ can be an index of the strength of segregation. 
Thus, our results suggest that an excessively large $\rho_{ls}$ deteriorates the segregation strength.
In a rotating drum with spherical particles, a large particle size ratio ($\rho_{ls} \sim$ 6) suppresses the radial segregation and band formation \cite{Thomas:RevSeg}. 
These results indicate that an excessively large $\rho_{ls}$ causes weaker or slower segregation, which supports our hypothesis. 
\\ \;
In this study, we determined rich pattern dynamics of segregation of a horizontally vibrated granular mixture in a quasi-two-dimensional container, which resembles the patterns in a reaction-diffusion system. 
In previous studies, coarsening bands were found in horizontally vibrated sub-monolayer granular mixtures, where the vibration amplitude was much smaller than the container size \cite{Mullin00:SegrelationPlate, Reis:SegregationCoarsening}. 
In contrast to our results, the bands were perpendicular to the vibration direction, except when the particle density ratio was very large \cite{Pihler:holvib, Inagaki:drumSizeDens}. 
Theoretical studies and simulations using the distinct element method predicted that the inertial effect and frictional force between the particle and container surface played a dominant role in band formation in sub-monolayer granular mixtures \cite{Pooley:modsegr, Fujii:SimuVib}. 
In our study, the phenomenological model suggests that complex pattern dynamics are derived from bistability of the granular flow. 
Thus, bulk rheology derived from particle-particle interactions could be a driving mechanism of band dynamics. 
\\ \;
The mechanism of band formation in our experiment could be related to that of the rotating drum. 
As we show in Figs. \ref{fig:fig3} (c) and \ref{fig:fig4} (b), surface flow regulated the dynamics of the band. 
In a rotating drum, the dynamics of the band bifurcate from the coarsening process to oscillatory motion through the traveling band when the strength of the surface flow is modulated by varying the fill levels \cite{Inagaki10:_Travelingwave, Inagaki15:Oscillatingwave}. 
At a high fill level, cross-sectional images of the granular mixture suggest that the shear from the surface flow drove radial and axial segregation \cite{Inagaki15:Oscillatingwave}. 
Experimental and theoretical studies have proposed that axial segregation in a rotating drum can be considered interfacial instability between the layer of large particles and that of small particles \cite{Taberlet:RadCore, Balista:_AxialCore}. 
Our observations from the side-view (Fig. \ref{fig:fig1} (c)) also implied that band formation in the horizontally vibrated granular mixture might behave as the interfacial instability. 
However, in the rotating drum, a mixture of spherical glass beads with different particle sizes formed clear bands, whereas we only observed subtle bands with the same combination of glass beads in our system (see supplementary information). 
Recently, it was reported that the size and density differences between the two types of granules dominantly determine the onset of axial segregation in a rotating drum \cite{Inagaki:drumSizeDens}, but band formation in our case had to be enhanced by the differences in frictional forces between two different particles.
\\ \;
In contrast to the granular Rayleigh Taylor instability \cite{Dortona20:_RT}, band formation occurs without a density difference. 
However, in our system, when we used granules with density differences, additional band dynamics, i.e., zig-zag droplets, were observed (see Supplementary Information). 
Thus, the granular Rayleigh Taylor instability can also modulate the band dynamics, as suggested in a rotating drum \cite{Inagaki:drumSizeDens}. 
However, the magnitude of the contribution of granular fluidity \cite{Dijksman:GraRheoHys}, density difference \cite{Dortona20:_RT}, and size differences to segregation dynamics remain unclear. 
Because our system enables us to study the time evolution of the inner structure of the granular mixture, further investigation of the interfaces between granular beds will help us to understand the detailed mechanism of the segregation pattern driven by the surface flow. 
\\ \;
{\bf Acknowledgments.}
\;
This work was supported by JSPS KAKENHI Grant Numbers JP19K14614, JP22K03468.

\end{document}